\documentclass[preprint,amssymb,amsmath,nobibnotes,nofootinbib]{revtex4}
\usepackage{graphicx}

\begin{document}
\title {Calculation of the Self Force using the Extended-Object Approach}

\author{Amos Ori and Eran Rosenthal \\
{\it
Department of Physics, Technion-Israel Institute of Technology, Haifa
32000, Israel}
}

\date{\today}

\begin{abstract}
We present here the extended-object approach for the explanation
and calculation of the self-force phenomenon (often called also
''radiation-reaction force''). In this approach, one considers a
charged extended object of a finite size $\epsilon$ that
accelerates in a nontrivial manner, and calculates the total force
exerted on it by the electromagnetic field (whose source is the
charged object itself). We show that at the limit $\epsilon
\rightarrow 0$ this overall electromagnetic field yields a
universal result, independent on the object's shape, which agrees
with the standard expression for the self force acting on a
point-like charge. This approach has already been considered by
many authors, but previous analyses ended up with expressions for
the total electromagnetic force that include $O(1/\epsilon )$
terms which do not have the form required by mass-renormalization.
(In the special case of a spherical charge distribution, this
$\propto 1/\epsilon $ term was found to be $4/3$ times larger than
the desired quantity.) We show here that this problem was
originated from a too naive definition of the notion of ''total
electromagnetic force'' used in previous analyses. Based on
energy-momentum conservation combined with proper relativistic
kinematics, we derive here the correct notion of total
electromagnetic force. This completely cures the problematic
$O(1/\epsilon )$ term, for any object's shape, and yields the
correct self force at the limit $\epsilon \rightarrow 0$. In
particular, for a spherical charge distribution, the above ''$4/3$
problem'' is resolved. We recently presented an outline of this
analysis \cite{OriRosenthal},
 focusing on the special case of a dumbbell-like charge distribution
 (i.e. two discrete point charges). Here we provide full detail of
 the analysis, and also extend it to more general charge distributions:
 extended objects with an arbitrary number of
 point charges, as well as continuously-charged objects.
\end{abstract}
\maketitle
\section{Introduction and summary}

When an electrically-charged particle accelerates (non-uniformly) in flat
spacetime, it exerts a force on itself. This force, known as the {\em self force} (or ''radiation-reaction force''), results from the particle's interaction
with its own electromagnetic field. Early investigations by
Abraham \cite{Abraham} and Lorentz \cite{Lorentz}
, in the case of non-relativistic motion, showed that the self
force is proportional to the time-derivative of the acceleration. Later,
Dirac \cite{Dirac} obtained the covariant relativistic expression for the self force:
\begin{equation}
f_{self}^{\mu }=\frac{2}{3}q^{2}(\dot{a}^{\mu }-a^{2}u^{\mu })\,\,,
\label{ALD}
\end{equation}
where $q$ is the electrical charge, $u^{\mu }$ and $a^{\mu }$ denote the four-velocity and
four-acceleration, respectively, an overdot denotes a proper-time
derivative, and $a^{2}\equiv a^{\mu }a_{\mu }$
\footnote{Throughout the paper we use the metric $\eta _{\mu \nu }=\text{diag}
(-1,1,1,1)$, and units where $c=1$.}.
Dirac derived this expression by considering the momentum flux through
a ''world-tube'' surrounding the particle's worldline, and demanding energy-momentum
conservation.

The fact that a particle can exert a force on itself is obviously
intriguing. One of the ways to make sense of this phenomenon is by
considering a charged, rigid, extended object of finite size $\epsilon $. A
model of a continuously-charged, finite-size object has the obvious
advantage that the electromagnetic field is everywhere regular, allowing (in
principle) an almost straightforward calculation of all electromagnetic
forces involved (this is of course not the case when a point-like charge is
considered, as the field is singular at the particle's location). On
physical grounds, one would expect that an extended object of a sufficiently
small size will behave like a point-like particle. (One should expect
finite-size correction terms, which may depend on the object's shape, but
one would hope these corrections would become negligible for a sufficiently
small object.)

For a finite-size extended object, each charge element exerts an
electromagnetic force on each other charge element. Then, roughly speaking,
the overall electromagnetic force that the charged object exerts on itself
is the sum of the contributions of all mutual forces between all pairs of
the object's charge elements. One might naively expect that this sum will
always vanish, by virtue of Newton's third law; Indeed, this would be the
case if the charged object were static: In this case, the sum of the Coulomb
forces vanishes for each pair separately. It turns out, however, that if the
object accelerates the sum of the contributions of all mutual forces does
{\em not} vanish (this has been established by many authors, as we discuss
below, and it is also demonstrated explicitly later in this paper). This
nonvanishing overall electromagnetic force does not conflict with the
momentum conservation law, because the electromagnetic field itself contains
a time-varying amount of momentum and energy; A non-vanishing overall
electromagnetic force acting on the object is thus a manifestation of a
momentum transfer between the charged object and the surrounding
electromagnetic field. In particular, the electromagnetic radiation field
carries energy and momentum away from the object to infinity (hence the name
''radiation-reaction force'').

Recognizing that the overall mutual electromagnetic force does not vanish,
one is tempted to identify this overall force with the notion of the self
force acting on a charged particle. Thus, one would hope that at the limit
were the object's size is taken to zero, a universal result (independent of
the object's size and shape) will be obtained, which will coincide with Eq. (%
\ref{ALD}). Many attempts have been made to derive this
extended-object total force. Two types of models have been considered:
objects that are continuously charged \cite{Lorentz,Schott,PanofPhil,GriffithsOwen}, and objects with a finite number of discrete charges \cite{GriffithsOwen}. The simplest model of a discretely-charged rigid
object is the ''dumbbell'', i.e. a fixed-length rod with two point charges
located at its two edges. The previous analyses of both the continuous and
discrete models revealed that indeed the overall electromagnetic force does
not generally vanish. But these analyses also indicated a fundamental
difficulty (which we shortly explain), that made it impossible to derive the
universal small-size limit of this force. The goal of this paper is to
provide a simple resolution to this difficulty.

Let $f_{sum}^{\mu }$ denote the sum of (or, in the continuous model, the
double-integral over) all mutual electromagnetic forces, acting on all
charge elements at a particular moment. (By ''particular moment'' we refer
here to a hypersurface of simultaneity in the particle's rest frame; see
below.) We would like to explore how $f_{sum}^{\mu }$ depends on $\epsilon $%
. For all types of electrically charged objects, the small-$\epsilon $ dependence of $%
f_{sum}^{\mu }$ is found to be of the form
\begin{equation}
f_{sum}^{\mu }=c^{\mu}_{-1}/\epsilon +c^{\mu}_{0}+O(\epsilon )\,.  \label{fsum0}
\end{equation}
The $O(\epsilon )$ term will not concern us here, as it vanishes at the
limit $\epsilon \rightarrow 0$. The coefficients $c^{\mu}_{0}$ and $c^{\mu}_{-1}$ depend
on the object's worldline, but are (by definition) independent of $\epsilon $.

The $O(\epsilon ^{-1})$ term is the problematic term, as it diverges at the limit of
interest, i.e. $\epsilon \rightarrow 0$. Obviously, the small-object limit
does not make sense if we do not know how to handle the problematic term $
c_{-1}/\epsilon $.

Now, there is a standard procedure of {\em mass-renormalization}, often used
for eliminating such $O(\epsilon ^{-1})$ terms. However, the very nature
of this procedure requires that the undesired $O(\epsilon ^{-1})$ term will
be of the form $-ca^{\mu }$, where $c$ is a parameter that is independent of
the time and the state of motion (though it may depend on the object's size
and shape): A force term of the form $a^{\mu }\cdot const$ can be dropped,
because it is experimentally indistinguishable from an inertial term in the
equation of motion (see next section). In order for the whole theory to make
sense (assuming that interaction energy equally contributes to the inertial
mass), the constant $c$ must be equal to the object's electrostatic energy,
which we denote $E_{es}$. (Recall that the latter scales like $\epsilon
^{-1} $.)

The problem is that, the term $c^{\mu}_{-1}/\epsilon $ is actually {\em not} equal
to $-E_{es}a^{\mu }$ --- and, furthermore, generally it is not in the form $%
-ca^{\mu }$, nor it is even in the direction of $a^{\mu }$. In the special
case of a spherical charge distribution, several authors found \cite{Lorentz,Schott,PanofPhil,GriffithsOwen} that the term $c^{\mu}_{-1}/\epsilon $ indeed takes the form $-ca^{\mu }$, but with $%
c=(4/3)E_{es}$ . This is the well-known ''$4/3$ problem''. In this special
case the mass-renormalization procedure still makes sense from the
operational point of view (because any force term of the form $-ca^{\mu }$
is experimentally indistinguishable from an inertial term), though the
logical consistency of the theory may be questioned. The situation is worse,
however, when the charge distribution is asymmetric. In such a situation one
generally finds that the problematic term $c^{\mu}_{-1}/\epsilon $ is not even in
the direction of $a^{\mu }$. Clearly, this type of divergent term cannot be
removed by mass renormalization. A simple demonstration of this situation
was given by Griffiths and Owen \cite{GriffithsOwen}, who considered a one-directional
motion of a dumbbell. They found that when the dumbbell is oriented
perpendicularly to the direction of motion, then $c^{\mu}_{-1}/\epsilon
=-E_{es}a^{\mu }$ as desired. However, if the dumbbell is co-directed with
the motion, then $c^{\mu}_{-1}/\epsilon =-ca^{\mu }$ with $c=2E_{es}$.
Furthermore, if the dumbbell is oriented in any other direction, the term $%
c_{-1}/\epsilon $ will not be co-directed with $a^{\mu }$. Clearly, in such
a generic situation the problematic term $c^{\mu}_{-1}/\epsilon $ cannot be
removed by mass renormalization. As a consequence, the limit $\epsilon
\rightarrow 0$ of $f_{sum}^{\mu }$ does not make a physical sense. We note
that this problem arises even if the object's motion is treated in a
fully-relativistic manner \cite{relativistic} --- as long as the quantity $f_{sum}^{\mu }$ is considered.

It should be noted that in the case of a spherical charge distribution there
is another ''$4/3$ problem'': When the object is in slow motion, the
electromagnetic-field momentum turns out to be $4/3$ times the
electromagnetic-field energy times the velocity \cite{Lorentz}. We may refer
to this problem as the ''inertial $4/3$ problem'' (as opposed to the
''mass-renormalization $4/3$ problem''). We shall not address this problem
here. The relation between these two ''$4/3$ problems'' is not completely
clear.\ Poincar\'{e} \cite{Poincare} introduced the non-electromagnetic internal stresses
in order to resolve the inertial $4/3$ problem. On the other hand, the
analysis presented below clearly indicates that no consideration of the
non-electromagnetic internal forces is required for solving the
mass-renormalization $4/3$ problem \cite{Review}.

In this paper we shall provide a simple and natural solution to the above
mass-renormalization problem. We shall show that the overall mutual
electromagnetic force is {\em not} the quantity $f_{sum}^{\mu }$ (i.e. the
naive sum or integration over all mutual forces); By employing simple
energy-momentum considerations we show that the overall mutual
electromagnetic force, which we denote $f_{mutual}^{\mu }$, is the sum (or
integral) over all mutual forces, each multiplied by a certain kinematic
factor representing the proper-time lapse of each charge element between two
''moments'' (i.e. between two neighboring hypersurfaces of simultaneity; see
next section). This kinematic factor is of the form $1+O(\epsilon )$; and
the $O(\epsilon )$ correction (when multiplying the mutual forces $\propto
\epsilon ^{-2}$) leads to a difference between $f_{sum}^{\mu }$ and $%
f_{mutual}^{\mu }$, proportional to $\epsilon ^{-1}$, which is exactly the
amount required to correct the problematic term $c^{\mu}_{-1}/\epsilon $. Namely,
when $f_{mutual}^{\mu }$ is expanded in powers of $\epsilon $, it takes a
form similar to Eq. (\ref{fsum0}), but with an $O(1/\epsilon )$ term which
is precisely of the form $-E_{es}a^{\mu }$. This $O(1/\epsilon )$ term is
naturally removed by mass renormalization.

After we have eliminated the problematic $O(\epsilon ^{-1})$ term in Eq.
(\ref{fsum0}), we are left with the regular term $c_{0}^{\mu }$. It is this
term which should yield the desired expression for the self force.
With the anticipation that the self force
should be universal, one would expect $c_{0}^{\mu }$ to depend only on the
object's total charge $q$, and not on the way it is distributed. In fact,
the very nature of the self-force phenomenon --- the force that a charge
exerts on itself --- suggests that the self force must be proportional to
$q^{2}$. For continuous charge distributions, the term $c_{0}^{\mu }$ is
indeed found to be $\propto q^{2}$ and it can be brought to the form (\ref
{ALD}). However, for discrete charge distributions $c_{0}^{\mu }$ is found
to depend on the charge distribution. This is best
demonstrated in the simplest discrete model, the dumbbell. In this case,
$c_{0}^{\mu }$ (like the mutual forces) is proportional to the product
$q_{+}q_{-}$, rather than to $q^{2}=(q_{+}+q_{-})^{2}$, where $q_{+}$ and
$q_{-}$ denote the two edge charges. This apparent inconsistency has an
obvious origin: The overall force exerted on the dumbbell by the
electromagnetic field includes not only the mutual forces between different
charges, but also the forces that each of the individual charges exerts
\textit{on itself} (to which we shall refer as the ``partial self force'',
to distinguish it from the ''overall self force'' acting on the dumbbell).
Obviously, it would be inconsistent to neglect these partial self forces: By
universality considerations, one may view each of the point charges as a
very small extended charged object; And, the result of our analysis, namely
$c_{0}^{\mu }\neq 0$, should apply to each of these individual charged
objects as well, therefore, these forces cannot be ignored. Note that the partial self forces do not depend on the other charges in the extended object, so they are by definition independent of $\epsilon $. Therefore they do not affect
the divergent term $O(\epsilon ^{-1})$, but merely add to the term
$c_{0}^{\mu }$. The need to include these partial self forces might appear
disturbing, as these quantities are initially unknown. However, basic
considerations imply that the self force acting on each charged object must
be proportional to the square of its charge. This observation provides us
with the required expression for the partial self forces (more precisely,
the relation of the latters to the overall self forces). The inclusion of
the partial self forces leads to a universal expression for the overall
electromagnetic force acting on the dumbbell (or on any other discrete
charge distribution), which is indeed proportional to the square of the
total charge as desired, and which coincides with Eq. (\ref{ALD}). We
further show in Sec.\ \ref{continuous} that for a continuous charge
distribution the contribution of the partial self forces vanishes.
Therefore, in the continuous case the quantity $c_{0}^{\mu }$ (i.e. the
properly weighted integral of the mutual electromagnetic forces) directly
yields the desired universal expression for the self force, Eq.\ (\ref{ALD}).

The overall mutual force $f_{mutual}^{\mu }$
 may naturally be viewed as the sum (or double-integral) of the contributions of
all {\em pairs} of charge element. The contribution of each such pair is the
sum of the two mutual forces, each weighted by the above mentioned kinematic
factor. In summing these two forces, the dominant $O(\epsilon ^{-2})$ term
always cancels out (leaving a weaker divergence $\propto \epsilon ^{-1}$
that is in turn handled by mass-renormalization). This leading-order
cancellation occurs for each pair separately, suggesting that the
fundamental element in any extended-object model is the single pair of
charges. Once the single-pair system is well understood, the analysis of any
charge distribution will follow quite immediately --- essentially by summing
(or double-integrating) over all pairs of charge elements. We shall
therefore start by analyzing the {\em dumbbell model}, i.e. a pair of
point-like charges separated by a fixed-length rod. Then we shall consider a
discrete system with an arbitrary number $N$ of charges. Then, taking the
infinitesimal limit (in which $N\rightarrow \infty $), we shall analyze the
case of continuous charge distribution. In all cases the object (and
the charge distribution) is regarded as rigid, and we allow it to move
(non-rotationally) along an arbitrary worldline. For both the discrete and
continuous cases, we obtain the same universal result: After calculating the
overall mutual electromagnetic force $f_{mutual}^{\mu }$, mass-renormalizing
it, and then taking the limit $\epsilon \rightarrow 0$, we recover the
desired expression (\ref{ALD}) for the self force.

We should mention here previous analyses which seemingly overcame
the mass-renormalization $4/3$ problem in the spherical case.
First, Fermi \cite{Fermi} carried out an extended-object analysis
of a different type: Instead of summing the contributions of all
mutual forces, he constructed an effective relativistic
Hamiltonian of a charged rigid body, and derived the equation of
motion from this Hamiltonian. It seems that no ''$4/3$ problem''
is encountered in this method. Later, Nodvick used a similar
method \cite{Nodvick} and obtained the correct expression for the
 self force (note, however, that these analyses
\cite{Fermi,Nodvick} only considered spherically symmetric
distributions, whereas we are treating here an arbitrary charge
distribution). Also, after this work was completed, we became
aware of a previous work by Pearle \cite{Pearle}, in which he
analyzed the case of a spherically-symmetric charged object. In
this analysis he took into account the above mentioned kinematic
weighting factor which expresses the proper-time lapse of each
charge element. Then, in a fairly complicated
calculation he obtained the correct $O(1/\epsilon )$ term, namely $%
-E_{es}a^{\mu }$, thereby overcoming the $4/3$ problem in the case of
spherical charge distribution. We believe that our analysis is simpler, more
transparent, and it is also much more general; In particular, the analysis
presented here resolves the mass-renormalization problem for {\em any} type
of charge distribution.

We point out that self force calculations were also carried out in
the context of quantum theory, see for example \cite{MS}. This
however does not diminish the need for a consistent classical
treatment of the self-force problem. Indeed in some situations the
classical framework is certainly the most natural one. Consider
for example the self force acting on an electrically-charged
satellite orbiting the earth. Obviously quantum effects are
irrelevant in this case, and it is therefore desired to treat this
situation from the purely classical point of view. An example of
much greater current interest is that of a compact object orbiting
a black hole of much larger mass, see e.g. \cite{letter}. The
compact object will gradually inspiral towards the black hole, due
to the {\it gravitational} self force. Again, there is no reason
to use the quantum theory in this case. Note that in this case one
must use the formalism of gravitational self force in {\it curved}
spacetime. Whereas the present paper only deals with flat space,
it appears that the extended-object approach can naturally be
extended to curved space as well. For example, Ori recently
applied the extended-object approach to the case of radial
free-fall in Schwarzschild spacetime \cite{Ori}, and used it to
calculate the regularization parameters \cite{letter}.

An outline of the analysis given here was published recently,
focusing on the case of dumbbell-like charge distribution
\cite{OriRosenthal}. Here we present the full calculations, and
also analyze in detail extended objects with an arbitrary number
of point charges, as well as continuously-charged extended objects
(these cases were only briefly mentioned in Ref.
\cite{OriRosenthal}).

In Sec. \ref{sec:dumbbell} we analyze the dumbbell model, i.e. the case of
two point-like charges. We first formulate the dumbbell's relativistic
kinematics. Then we calculate the mutual forces, obtain their sum $%
f_{sum}^{\mu }$, and (following Griffiths and Owen \cite{GriffithsOwen}) demonstrate the
severe mass-renormalization directionality problem discussed above. Then we
use energy-momentum considerations to construct the correct expression for
the overall mutual electromagnetic force $f_{mutual}^{\mu }$. We show that
the latter is free of the mass-renormalization problem. In Sec. \ref{Nbody}
we extend the analysis to a system with an arbitrary number of point-like
charges. Finally, in Sec. \ref{continuous} we consider the case of a
continuous charge distribution. In all three cases we obtain, at the limit $%
\epsilon \rightarrow 0$, the universal result (\ref{ALD}).

\section{A charged Dumbbell}
\label{sec:dumbbell}
\subsection{The general approach}

We consider a dumbbell made of a rigid rod with two point charges located at
its two edges. The forces acting on the dumbbell (or on its parts) may be
schematically divided into several types:

\begin{itemize}
\item  Electromagnetic forces: the forces exerted on the two charges by the
electromagnetic field they produce;

\item  The ''other internal forces'': the inter-atomic (or ''elastic'')
forces that are responsible for the dumbbell's rigidity;

\item  External forces: forces exerted on the dumbbell by external fields.%
\footnote{%
Throughout this paper, by ''electromagnetic forces'' we shall always refer
to the forces associated with the interaction of the two charges with the
electromagnetic field they themselves produce, and {\em not} to external
forces (or the ''other internal forces'') even if the latters are of
electromagnetic origin. Namely, our terminology is based on the simplified
picture according to which the external forces and the ''other internal
forces'' are non-electromagnetic. However, this presumption is only made for
simplifying the terminology, and the analysis below
is valid even if these forces are of electromagnetic character.}
\end{itemize}

The electromagnetic forces acting on the two charges are divided into two
types: (i) Mutual electromagnetic forces, i.e. forces that one charge exerts
on the other one [more precisely, it is the force that the electromagnetic
field produced by one charge (in the sense of the retarded Lienard-Wiechert
solution) exerts on the other charge]; and (ii) the self forces that each of
the two charges exerts on itself, to which we shall refer as the ''partial
self forces'' (to be distinguished from the overall self force acting on the
dumbbell). The justification and necessity of including the partial self
forces in our analysis is discussed below; but two remarks should be made
already at this stage: First, the partial self forces are not
relevant to the mass-renormalization problem, as they only affect the term $c^{\mu}_{0}$
(above), not the problematic term $c^{\mu}_{1}/\epsilon $. Second, these partial
self forces are very significant for a system of two charges (they
contribute at least as much as the mutual forces do), but they become less
important in a system including a large number $N$ of point charges
(assuming that the magnitude of the individual charges scales line $1/N$).
This is because the number of mutual forces scales like $N^{2}$, whereas the number
of partial self forces scales like $N$. Most importantly, the contribution
of partial self forces vanishes at the continuum limit, as we discuss in
Sec.\ \ref{continuous}.

In Newtonian theory it is usually presumed that the sum of any pair of
mutual forces will always vanish; However, when electromagnetic interactions
are concerned, this presumption does not hold. Its failure may be attributed
to the long range of the electrodynamical interaction between two charges.
It is this long range which is responsible for the electromagnetic radiative
phenomena (which transport energy and momentum away from the interacting
charges). On the other hand, the non-electromagnetic internal forces are
assumed here to be of ''short range''.\footnote{%
For our discussion it is sufficient to assume that the range of the ''other
internal forces'' is small compared to the dumbbell's length. This
assumption perfectly holds for e.g. the inter-atomic forces that are
responsible for the rod's rigidity. It is also sufficient to assume that this range is small compared to
the time scales charactering  the world line $z^\mu(\tau)$. e.g. $1/a$, $a/\dot{a}$ etc.} Hence, it will be assumed that upon summation these forces will always cancel out (except for a ``mass-renormalization like'' term, which is the interaction energy associated with these forces, multiplied by the four-acceleration). For this reason, the
non-external forces that are relevant to the calculations below are only the
electromagnetic ones, namely, (i) the two mutual forces between the two
charges, and (ii) the two ''partial self forces''.

\subsection{Dumbbell's structure and kinematics}

The dumbbell consists of two point charges situated at the edges of a rigid
rod of a proper length $2\epsilon $. We shall assume that $\epsilon $ is
small compared to $1/a$, where $a$ denotes the norm of the acceleration
vector. Throughout this section we shall use the subscripts ''$+$'' or ''$-$%
'' to denote the quantities associated with the two dumbbell's edges. The
two electric charges are therefore denoted $q_{+}$ and $q_{-}$,
respectively, and the total charge is $q\equiv q_{+}+q_{-}$.
We do {\it not} require the two charges to be equal.
We assume that $\epsilon $ is
time-independent and that the dumbbell moves in a non-rotational manner (see
below).

\begin{figure}
\includegraphics{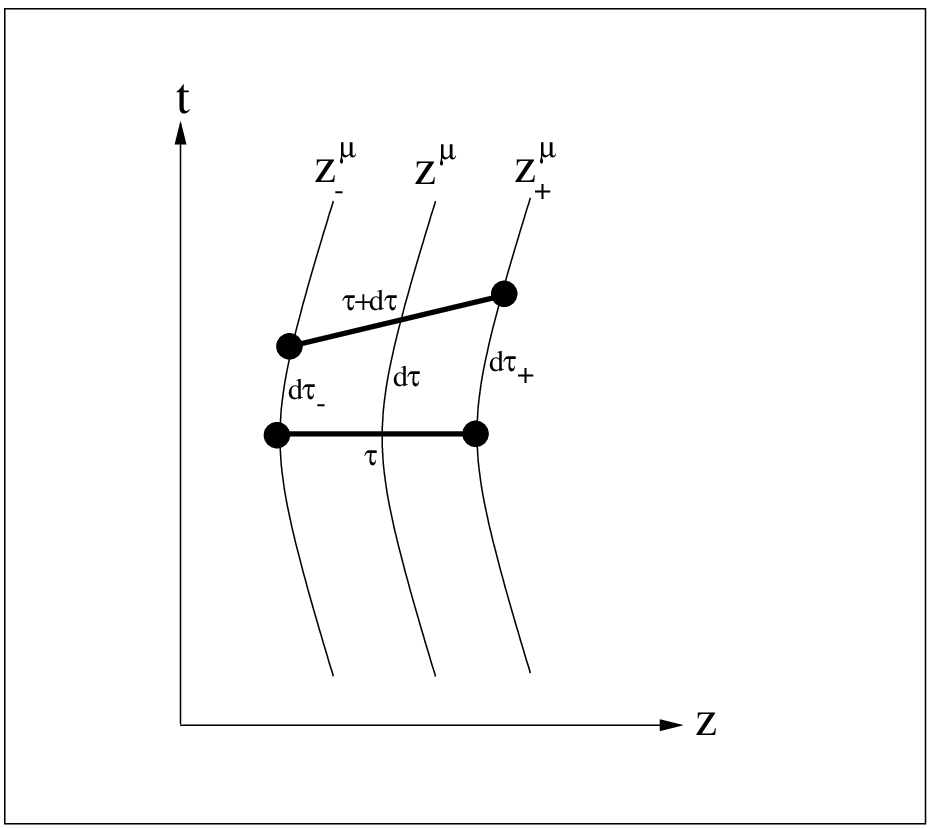}
\caption{ A spacetime diagram describing the dumbbell's
kinematics. t is the time coordinate (in some inertial reference
frame), and z schematically represents a spatial coordinate. The
dumbbell is represented by a straight bold line, with the black
points representing the two edge points $z_\pm^\mu$. Two such bold
lines are shown, representing the dumbbell's location in spacetime
at two moments separated by an infinitesimal time interval
$d\tau$.
 The three thin solid lines are
the worldlines of the central point $z^\mu$ and the two edge
points $z_\pm^\mu$. \label{fig1}}
\end{figure}

We take the dumbbell's central point (i.e. half the way between the two
edges) to represent the dumbbell's motion. The worldline of this
representative point is denoted $z^{\mu }(\tau )$, where $\tau $ is the
proper time along the central worldline. The four-velocity and
four-acceleration of the central worldline are defined in the usual manner, $%
u^{\mu }\equiv \dot{z}^{\mu }$ and $a^{\mu }\equiv \dot{u}^{\mu }$, where an
overdot denotes differentiation with respect to $\tau $. We denote the two
rod's edges by $z_{+}^{\mu }(\tau )$ and $z_{-}^{\mu }(\tau )$. At any given
moment (by ''moment'' we mean here a hypersurface of simultaneity in the
momentary rest frame of the central point) the two rod's edges are located at
(see Fig.\ \ref{fig1})
\begin{equation}
z_{\pm }^{\mu }(\tau )\equiv z^{\mu }(\tau )\pm \epsilon w^{\mu }(\tau )\ ,
\label{zpm}
\end{equation}
where $w^{\mu }(\tau )$ is a unit spatial vector, satisfying
\begin{equation}
\begin{array}{ccc}
w_{\mu }w^{\mu }=1 & , & w^{\mu }u_{\mu }=0\ .
\end{array}
\label{wdef}
\end{equation}

The time evolution of $w^{\mu }$ is subjected to two restrictions: First, as
a unit vector, its norm is time-independent (corresponding to a rod of fixed
length). Second, it is non-rotating in the momentary rest frame of $z^{\mu
}(\tau )$. These restrictions correspond to a Fermi-Walker transport
\cite{MTW} of $w^{\mu }$ along $z^{\mu }(\tau )$, given by
\begin{equation}
\dot{w}^{\mu }=(u^{\mu }a^{\nu }-u^{\nu }a^{\mu })w_{\nu }=u^{\mu }a_{w}\ ,
\label{fw}
\end{equation}
where the scalar $a_{w}$ denotes the projection of $a^{\mu }$ on the rod's
direction: $a_{w}\equiv a_{\lambda }w^{\lambda }$. This transport rule
guarantees that if Eq.\ (\ref{wdef}) is initially satisfied, it will hold at
all subsequent times.

Next, we calculate the four-velocities and accelerations of the dumbbell's
edges. We denote the proper times along the worldlines of the two rod's
edges $z_{\pm }^{\mu }$ by $\tau _{\pm }$, respectively. Note that generally
$\tau _{+}$ and $\tau _{-}$ differ from $\tau $ (and from each other). The
four-velocities of the two charges are defined in the usual manner, $u_{\pm
}^{\mu }\equiv dz_{\pm }^{\mu }/d\tau _{\pm }$. Differentiating Eq.\ (\ref
{zpm}), with respect to $\tau _{\pm }$, we obtain
\begin{equation}
u_{\pm }^{\mu }=(u^{\mu }\pm \epsilon \dot{w}^{\mu })\frac{d\tau }{d\tau
_{\pm }}=\left[ (1\pm \epsilon a_{w})\frac{d\tau }{d\tau _{\pm }}\right]
u^{\mu }\ .  \label{vedge}
\end{equation}
Taking the norm of the two sides of this equation, recalling that both $%
u^{\mu }$ and $u_{\pm }^{\mu }$ are of unit norm, we find that the term in
squared brackets is just unity, namely

\begin{equation}
\frac{d\tau _{\pm }}{d\tau }=1\pm \epsilon a_{w}\ .  \label{taupmtau}
\end{equation}
It now immediately follows that
\begin{equation}
u_{\pm }^{\mu }=u^{\mu }\ .  \label{vedge2}
\end{equation}
These are the two key features of the rod's kinematics.

Eq.\ (\ref{vedge2}) indicates that in the rest frame of the dumbbell's
central point, the two edges (and similarly any other point on the dumbbell)
are at rest as well. We can therefore identify this reference frame as the
rest frame of the entire dumbbell. Since at any moment there exists a
reference frame in which the entire dumbbell is momentarily at rest, it is
justified to view this type of motion as a rigid motion.

We denote by $a_{\pm }^{\mu }$ the four-accelerations of the two edge
points, namely $a_{\pm }^{\mu }=du_{\pm }^{\mu }/d\tau _{\pm }$. From Eqs.\ (%
\ref{taupmtau}) and (\ref{vedge2}) it immediately follows that
\begin{equation}
a_{\pm }^{\mu }=\frac{a^{\mu }}{1\pm \epsilon a_{w}}\ .  \label{aedge}
\end{equation}

\subsection{Mutual forces}

At the heart of the dumbbell's model are the mutual forces acting between the
two charges. To determine these forces, we need an expression for the retarded
electromagnetic field tensor $F_{\mu \nu }$ that a single point charge $q$ moving
on an arbitrary worldline $z^{\mu }(\tau )$ produces at a nearby point $%
z^{\mu }+\hat{\epsilon}\hat{w}^{\mu }$, where $\hat{\epsilon}$ is a small
positive number ($\hat{\epsilon}=2\epsilon $), and $\hat{w}^{\mu }$ is a
unit spatial vector satisfying $\hat{w}^{\mu }\hat{w}_{\mu }=1,\hat{w}^{\mu
}u_{\mu }=0$. Later we shall apply the limit $\epsilon \rightarrow 0$, and
therefore we shall only need an expression for $F_{\mu \nu }$ valid up to
zero order in $\hat{\epsilon}$. Such an expression was derived by Dirac \cite{Dirac}:
\begin{eqnarray}
F_{\mu \nu } &\cong &\frac{q}{\sqrt{(1+\hat{\epsilon}a_{\hat{w}})}}\left[ (%
\frac{u_{\mu }\hat{w}_{\nu }}{\hat{\epsilon}^{2}}+\frac{a_{\mu }u_{\nu }}{2%
\hat{\epsilon}}+\frac{a^{2}u_{\mu }\hat{w}_{\nu }}{8}-\frac{\dot{a}_{\mu }%
\hat{w}_{\nu }}{2}-\frac{a_{\hat{w}}a_{\mu }u_{\nu }}{2}-\frac{2}{3}\dot{a}%
_{\mu }u_{\nu })-(\mu \leftrightarrow \nu )\right]  \nonumber \\
&\cong &q\ \left[ (\frac{u_{\mu }\hat{w}_{\nu }}{\hat{\epsilon}^{2}}+\frac{%
a_{\mu }u_{\nu }-a_{\hat{w}}u_{\mu }\hat{w}_{\nu }}{2\hat{\epsilon}}-\frac{2%
}{3}\dot{a}_{\mu }u_{\nu }+\hat{Z}_{\mu \nu })-(\mu \leftrightarrow \nu
)\right] ,  \label{fmunu}
\end{eqnarray}
where $a_{\hat{w}}\equiv a_{\lambda }\hat{w}^{\lambda }$,
\begin{equation}
\hat{Z}_{\mu \nu }\equiv \frac{a^{2}u_{\mu }\hat{w}_{\nu }}{8}-\frac{\dot{a}%
_{\mu }\hat{w}_{\nu }}{2}+\frac{3a_{\hat{w}}^{2}u_{\mu }\hat{w}_{\nu }}{8}%
-\frac{3a_{\hat{w}}a_{\mu }u_{\nu }}{4}\,,  \label{Z}
\end{equation}
$a^{2}\equiv a_{\mu }a^{\mu }$, and throughout this paper the ''$\cong $''
symbol denotes an equality up to $O(\epsilon )$ correction terms. $\hat{Z}%
_{\mu \nu }$ is the collection of all terms that are proportional to $\hat{%
\epsilon}^{0}$ and to an odd power of $\hat{w}$. Such terms will cancel out
when summing the contributions of the two charges (see below). The
electromagnetic field $F_{+}^{\mu \nu }$ that the charge $q_{-}$ produces at
the location of charge $q_{+}$ is obtained by substituting in Eqs.\ (\ref
{fmunu},\ref{Z}) $q\rightarrow q_{-}$, $a\rightarrow a_{-}$, $\dot{a}%
\rightarrow da_{-}/d\tau _{-}$, $\hat{w}^{\mu }\rightarrow w^{\mu }$, $a_{%
\hat{w}}\rightarrow a_{-}^{\lambda }w_{\lambda }$, and $\hat{\epsilon}
\rightarrow 2\epsilon \ $(the four-velocity is unchanged, as $u_{\pm }^{\mu
}=u^{\mu }$). The electromagnetic field $F_{-}^{\mu \nu }$ that the charge $%
q_{+}$ produces at the location of the charge $q_{-}$ is obtained in a
similar manner, by substituting in these equations $q\rightarrow q_{+}$, $%
a\rightarrow a_{+}$, $\dot{a}\rightarrow da_{+}/d\tau _{+}$, $\hat{w}^{\mu
}\rightarrow -w^{\mu }$, $a_{\hat{w}}\rightarrow -a_{+}^{\lambda }w_{\lambda
}$, and $\hat{\epsilon}\rightarrow 2\epsilon $. Let us denote by $f_{+}^{\mu
}$ ($f_{-}^{\mu }$) the Lorentz force that the charge ''$-$'' (''$+$'')
exerts on the other charge ''$+$'' (''$-$''):
\[
f_{\pm }^{\mu }=q_{\pm }F_{\pm }^{\mu \nu }u_{\nu }\,.
\]
By virtue of Eq. (\ref{fmunu}) this becomes
\begin{equation}
f_{\pm }^{\mu }\cong q_{+}q_{-}u_{\nu }\left[ (\pm \frac{u^{\mu }w^{\nu }}{%
4\epsilon ^{2}}+\frac{a_{\mp }^{\mu }u^{\nu }-(a_{\mp }^{\lambda }w_{\lambda
})u^{\mu }w^{\nu }}{4\epsilon }-\frac{2}{3}\dot{a}_{\mp }^{\mu }u^{\nu }\pm
Z_{\pm }^{\mu \nu })-(\mu \leftrightarrow \nu )\right] \ \,,  \label{11}
\end{equation}
where
\[
Z_{\pm }^{\mu \nu }\equiv \frac{a_{\mp }^{2}u^{\mu }w^{\nu }}{8}-\frac{\dot{a%
}_{\mp }^{\mu }w^{\nu }}{2}+\frac{3(a_{\mp }^{\lambda }w_{\lambda
})^{2}u^{\mu }w^{\nu }}{8}-\frac{3(a_{\mp }^{\lambda }w_{\lambda })a_{\mp }^{\mu
}u^{\nu }}{4}\,,
\]
and $\dot{a}_{\mp }^{\mu }\equiv da_{\mp }^{\mu }/d\tau _{\mp }$. Next we
re-express $f_{\pm }^{\mu }$ in terms of the acceleration $a^{\mu }$ and
proper time $\tau $ of the central point (rather than those of the source
charges). To this end we use Eq. (\ref{aedge}), and expand $a_{\mp }^{\mu }$
in $\epsilon $. Since the acceleration does not appear in the $O(\epsilon
^{-2})$ term, it is sufficient to carry out this expansion up to first order
in $\epsilon $:
\[
a_{\mp }^{\mu }=a^{\mu }\left( 1\pm \epsilon a_{w}\right) +O(\epsilon
^{2})\,.
\]
Note that $\dot{a}_{\mp }^{\mu }$ only appears in the $O(\epsilon ^{0})$
term, hence it can be replaced by $\dot{a}^{\mu }$. [The same holds for all
factors $a_{\mp }^{\mu }$ and $a_{\mp }^{2}$ that appear in the $O(\epsilon
^{0})$ term.] We find
\begin{equation}
f_{\pm }^{\mu }\cong q_{+}q_{-}u_{\nu }\left[ (\pm \frac{u^{\mu }w^{\nu }}{
4\epsilon ^{2}}+\frac{a^{\mu }u^{\nu }-a_{w}u^{\mu }w^{\nu }}{4\epsilon }-%
\frac{2}{3}\dot{a}^{\mu }u^{\nu }\pm Z^{\mu \nu })-(\mu \leftrightarrow \nu
)\right] \,,  \label{12}
\end{equation}
where
\[
Z^{\mu \nu }\equiv Z_{\pm }^{\mu \nu }(a_{\mp }^{\mu }\rightarrow a^{\mu })+%
\frac{a_{w}a^{\mu }u^{\nu }-a_{w}^{2}u^{\mu }w^{\nu }}{4}\,.
\]
Note that $Z^{\mu \nu }$ is $O(\epsilon ^{0})$, and is the same for the two
charges. Recalling that $u^{\nu }u_{\nu }=-1$, $w^{\nu }u_{\nu }=a^{\nu
}u_{\nu }=0$, and $\dot{a}^{\nu }u_{\nu }=-a^{2}$ (the latter identity is
obtained by differentiating $a^{\nu }u_{\nu }=0$), we find

\begin{equation}
f_{\pm }^{\mu }\cong q_{+}q_{-}\ \left[ \pm \frac{w^{\mu }}{4\epsilon ^{2}}-%
\frac{a^{\mu }+w^{\mu }a_{w}}{4\epsilon }+\frac{2}{3}(\dot{a}^{\mu
}-a^{2}u^{\mu })\pm Z^{\mu }\right] \,,  \label{mutf}
\end{equation}
where $Z^{\mu }\equiv u_{\nu }(Z^{\mu \nu }-Z^{\nu \mu })$.

\subsection{Naive sum of the mutual forces}

Next we calculate the sum of the two mutual forces, i.e. the quantity $%
f_{sum}^{\mu }$:
\begin{equation}
f_{sum}^{\mu }\equiv f_{+}^{\mu }+f_{-}^{\mu }\cong -\frac{q_{+}q_{-}}{%
2\epsilon }(a^{\mu }+w^{\mu }a_{w})+\frac{4}{3}q_{+}q_{-}(\dot{a}^{\mu
}-a^{2}u^{\mu })\,.  \label{fsum}
\end{equation}
This quantity would be the simplest candidate for the dumbbell's self force;
However, as already discussed in the previous section, it suffers from a
serious problem: The first term on
the right-hand side is proportional to $%
1/\epsilon $, and hence diverges at the limit of interest, $\epsilon
\rightarrow 0$. The usual way to eliminate
 such an undesired $O(\epsilon^{-1})$ term is by the procedure of {\em mass renormalization} (see below);
However, from the very nature of this procedure, it will only be applicable
if the term to be removed is of the form $a^{\mu }\cdot const$ (a constant
that scales like $1/\epsilon $). Instead, in Eq. (\ref{fsum}) the term $%
a^{\mu }+w^{\mu }a_{w}$ is orientation-dependent. Furthermore, this term is
{\em not} co-directed with $a^{\mu }$. This difficulty was observed by
Griffiths and Owen \cite{GriffithsOwen}. \footnote{%
When a spherical charge distribution is considered, after integrating $%
f_{sum}^{\mu }$ over the charge distribution, one obtains an overall force
which is co-directed with $a^{\mu }$, due to the symmetry. In this case the
only imprint of the problem in the $O(1/\epsilon )$ term of $f_{sum}^{\mu }$
is the ''$4/3$ problem''. However, for generic non-symmetric distributions,
the integrated force will not be co-directed with $a^{\mu }$.} [Note that
adding the two ''partial self forces'' would not change this situation, as
it does not affect the $O(\epsilon ^{-1})$ term -- see below.]

\subsection{Energy-momentum balance}

The above pathology of the $O(1/\epsilon )$ term clearly indicates that $%
f_{sum}^{\mu }$ is not a valid candidate for the dumbbell's self force. The
reason is that, $f_{sum}^{\mu }$ does not correctly represent the overall
mutual force. To understand the reason for this, we shall now employ simple
considerations of energy-momentum conservation. These considerations will
indicate the
appropriate way to sum the two mutual forces, in
order to obtain the correct expression for the overall mutual force.

Let us denote the total dumbbell's four-momentum, at a given
moment $\tau $, by $p^{\mu }(\tau )$. This quantity is to be
obtained by integrating the appropriate components of the
dumbbell's stress-energy tensor over the hypersurface of
simultaneity, which we denote $\sigma $. Recalling that $u^{\nu}$
is normal to $\sigma $, we may write this integral as
\begin{equation}
p^{\mu }\equiv -\int_{\sigma }T_{(dumb)}^{\mu \nu }u_{\nu }d^{3}\sigma \,.
\label{defmom}
\end{equation}
Here $d^{3}\sigma $ is a volume element, and $T_{(dumb)}^{\mu \nu }$ denotes
the dumbbell's stress-energy tensor, {\em not including} the electromagnetic
field. The integration is performed over the entire volume of the dumbbell
(the integrand vanishes off the dumbbell).

It is worth emphasizing two points here: First, the integration is carried
out over a {\em hypersurface of simultaneity}, defined at each moment by the
dumbbell's motion, and {\em not} over a hypersurface $t=const$ of some fixed
Lorentz frame. This is the natural covariant way to define the
time-dependent four-momentum of a rigid body. \footnote{%
Note that an integration over a hypersurfaces $t=const$
of the Lorentz frame in use would produce a quantity $p^{\mu }(\tau )$ that transforms in
a complicated, non-covariant manner in a Lorentz transformation. On the
other hand, our $p^{\mu }(\tau )$ (defined by integration over the
hypersurface of simultaneity) transforms like a four-vector, as desired.}.
Second, we choose not to include the electromagnetic stress-energy tensor in
$p^{\mu }$, because the electromagnetic contribution is not well localized:
It is partly scattered throughout the space in the form of electromagnetic
waves. The non-electromagnetic part, however, is by assumption
well-localized, and hence monitoring $p^{\mu }(\tau )$ will provide us with
the desired information concerning the dumbbell's motion. Note that the
external field (i.e. the above mentioned ''external force'') is also not
included in $T_{(dumb)}^{\mu \nu }$.

From energy-momentum conservation it follows that $p^{\mu }(\tau )$ will
only change due to external forces acting on the dumbbell (if such exist),
and due to energy-momentum exchange between the dumbbell and the
electromagnetic field. The electromagnetic energy-momentum exchange is
manifested by the electromagnetic forces acting on the two charges. In an
infinitesimal time interval $d\tau $, the change in $p^{\mu }(\tau )$ will
be given by
\begin{equation}
dp^{\mu }=dp_{+}^{\mu }+dp_{-}^{\mu }+dp_{ext}^{\mu }\,,  \label{dpt}
\end{equation}
where $dp_{ext}^{\mu }$ is the contribution of the external force, and $%
dp_{\pm }^{\mu }$ denote the contributions from the electromagnetic forces
acting on the two charges \footnote{%
By ''electromagnetic forces'' we refer here to the forces exerted on the $%
\pm $ charges by the electromagnetic fields produced by these two charges,
as explained above.}. Let us denote these electromagnetic forces by $%
f_{(em)\pm }^{\mu }$. As discussed above, $f_{(em)\pm }^{\mu }$ includes
both the mutual electromagnetic force $f_{\pm }^{\mu }$, and the partial
self force acting on the $\pm $ charge, which we denote $\hat{f}_{\pm }^{\mu
}$:
\begin{equation}
f_{(em)\pm }^{\mu }=f_{\pm }^{\mu }+\hat{f}_{\pm }^{\mu }\,.  \label{tot}
\end{equation}
Note that simple consistency considerations require us to include the
partial self forces in the analysis: Our calculation shows (as many previous
analyses did) that there is a nonvanishing self force acting on a charged
object (the dumbbell, in our specific model); This force is found to be
universal (at the limit of small $\epsilon $), namely it is independent
of the object's size and orientation. It must therefore apply to {\em any}
sufficiently-small charged object -- and, in particular, to the two point
charges $q_{+}$ and $q_{-}$. Later we shall employ a simple argument to
quantitatively relate the two partial self forces $\hat{f}_{\pm }^{\mu }$ to
the overall self force acting on the dumbbell. (It should be emphasized that
the calculation below yields a nonvanishing overall self force even if one
does not take into account the partial self forces; Nevertheless the
resultant expression for the self force would be incorrect in such a case,
due to the inconsistency.) Note that the need for adding the partial self
forces is also made obvious from the following observation: Without the
partial self forces, the overall mutual electromagnetic force is
proportional to the product $q_{+}q_{-}$, whereas the overall self force of
the dumbbell (like that of any charged particle) must be proportional to $%
q^{2}=(q_{+}+q_{-})^{2}$. Adding the partial self forces
compensates for this difference exactly, as we show below.

Let us now calculate $dp_{+}^{\mu }$, the energy-momentum exchange of the ''$%
+$'' charge with the electromagnetic field, between the two hypersurfaces of
simultaneity $\tau $ and $\tau +d\tau $. An observer located at the ''$+$''
charge will measure a proper-time interval $d\tau _{+}$ between these two
hypersurfaces. Therefore, the amount of electromagnetic energy-momentum
transfer is $dp_{+}^{\mu }=f_{(em)+}^{\mu }d\tau _{+}$. Similar
considerations will apply of course to the other charge ''$-$''; therefore,

\begin{equation}
dp_{\pm }^{\mu }=f_{(em)\pm }^{\mu }d\tau _{\pm }\,.  \label{dp}
\end{equation}

Combining equations (\ref{dpt}), (\ref{dp}) and (\ref{tot}), we obtain

\begin{equation}
dp^{\mu }=(f_{+}^{\mu }+\hat{f}_{+}^{\mu })d\tau _{+}+(f_{-}^{\mu }+\hat{f}
_{-}^{\mu })d\tau _{-}+dp_{ext}^{\mu }\,.  \label{dp1}
\end{equation}
Defining the overall force acting on the system to be $f^{\mu }\equiv
dp^{\mu }/d\tau $, we find
\begin{equation}
f^{\mu }\cong \left[ \left\{ f_{+}^{\mu }\frac{d\tau _{+}}{d\tau }
+f_{-}^{\mu }\frac{d\tau _{-}}{d\tau }\right\} +(\hat{f}_{+}^{\mu }+\hat{f}
_{-}^{\mu })\right] +f_{ext}^{\mu }\,.  \label{force}
\end{equation}
Note that since the external force is presumably regular (i.e. it is
well-behaved at the limit of small $\epsilon $), and $d\tau _{\pm }/d\tau
\rightarrow 1$ at the limit $\epsilon \rightarrow 0$, we can simply take $%
dp_{ext}^{\mu }\cong f_{ext}^{\mu }d\tau $. For the same reason, since the
partial self forces are presumably regular too, we can ignore the factors $%
d\tau _{\pm }/d\tau $ multiplying $\hat{f}_{\pm }^{\mu }$. It is only the
mutual force $f_{\pm }^{\mu }$, which includes negative powers of $%
\epsilon $, that requires one to make the distinction between $d\tau $
and $d\tau _{\pm }$.

The overall mutual electromagnetic force $f_{mutual}^{\mu }$ is the term in
curly brackets in Eq. (\ref{force}):
\begin{equation}
f_{mutual}^{\mu }=f_{+}^{\mu }\frac{d\tau _{+}}{d\tau }+f_{-}^{\mu }\frac{%
d\tau _{-}}{d\tau }\,.  \label{dambmut0}
\end{equation}
Using Eqs. (\ref{taupmtau}) and (\ref{mutf}), and again neglecting terms
that vanish as $\epsilon \rightarrow 0$, we find
\[
f_{\pm }^{\mu }\frac{d\tau _{\pm }}{d\tau }=(1\pm \epsilon a_{w})f_{\pm
}^{\mu }\cong q_{+}q_{-}\ \left[ \pm \frac{w^{\mu }}{4\epsilon ^{2}}-\frac{%
a^{\mu }}{4\epsilon }+\frac{2}{3}(\dot{a}^{\mu }-a^{2}u^{\mu })\pm \tilde{Z}
^{\mu }\right] \,,
\]
where
\[
\tilde{Z}^{\mu }=Z^{\mu }-\frac{a_{w}(a^{\mu }+w^{\mu }a_{w})}{4}\,.
\]
It now follows that

\begin{equation}
f_{mutual}^{\mu }\cong -\frac{q_{+}q_{-}}{2\epsilon }a^{\mu }+\frac{4}{3}
q_{+}q_{-}(\dot{a}^{\mu }-a^{2}u^{\mu })\,.  \label{dambmut}
\end{equation}

The overall electromagnetic contribution to the total force $f^{\mu }$
acting on the dumbbell (not including the external force) is the term in
squared brackets in Eq. (\ref{force}), i.e. the sum of $f_{mutual}^{\mu }$
and the two partial self forces. We shall refer to it as the ''bare self
force'' (because subsequently we shall apply to it the mass-renormalization
procedure, to obtain the ''renormalized self force''), and denote it $%
f_{bare}^{\mu }$. It is given by

\begin{eqnarray}
f_{bare}^{\mu } &=&f_{mutual}^{\mu }+(\hat{f}_{+}^{\mu }+\hat{f}_{-}^{\mu })
\nonumber \\
&=&-\frac{q_{+}q_{-}}{2\epsilon }a^{\mu }+\frac{4}{3}q_{+}q_{-}(\dot{a}^{\mu
}-a^{2}u^{\mu })+(\hat{f}_{+}^{\mu }+\hat{f}_{-}^{\mu })+O(\epsilon )\,.
\label{fbare}
\end{eqnarray}

\subsection{Mass-renormalization and the renormalized self force}

In Eq. (\ref{fbare}) [like in Eq. (\ref{dambmut})] the $O(1/\epsilon )$ term
has the desired form $-E_{es}a^{\mu }$, where $E_{es}$ is the dumbbell's
electrostatic energy (at rest):

\[
E_{es}\equiv q_{+}q_{-}/2\epsilon \,.
\]
This is exactly the type of $O(1/\epsilon )$ term that is cured by mass
renormalization, as we now briefly discuss.

The expression for the self force is to be used for predicting the
dumbbell's motion, through an equation of motion of the form $m_{bare}a^{\mu
}=f^{\mu }$, where $f^{\mu }$ refers to the total force acting on the
dumbbell, i.e. $f^{\mu }=f_{bare}^{\mu }+f_{ext}^{\mu }$. (Below we shall
further discuss the justification to this equation of motion.) Similarly, $%
m_{bare}$ refers to the so-called ''bare mass'', i.e. the total dumbbell's
energy (in the momentary rest frame) {\em not including the
electromagnetic/electrostatic interaction energy}. We now add the term $%
E_{es}a^{\mu }$ to both sides of the equation of motion. Defining the
''renormalized mass'' $m_{ren}$ and ''renormalized self force'' $%
f_{ren}^{\mu }$ by
\begin{equation}
m_{ren}\equiv m_{bare}+E_{es}\,,\,\,\,\,f_{ren}^{\mu }\equiv f_{bare}^{\mu
}+E_{es}a^{\mu }\,,  \label{massren}
\end{equation}
the equation of motion now takes the form
\[
m_{ren}a^{\mu }=f_{ren}^{\mu }+f_{ext}^{\mu }\,.
\]
This is the ''renormalized equation of motion''. Note that $m_{ren}$ is
nothing but the total dumbbell's energy (including the electrostatic
interaction) while at rest. This is in fact the measured physical mass of
the dumbbell. To simplify the notation, we shall hereafter omit the suffix
''ren'', denoting the renormalized mass by $m$ and the ''renormalized self
force'' by $f_{self}^{\mu }$. The equation of motion now reads
\[
ma^{\mu }=f_{self}^{\mu }+f_{ext}^{\mu }\,,
\]
where
\begin{equation}
f_{self}^{\mu }\equiv f_{bare}^{\mu }+E_{es}a^{\mu }=\frac{4}{3}q_{+}q_{-}(%
\dot{a}^{\mu }-a^{2}u^{\mu })+(\hat{f}_{+}^{\mu }+\hat{f}_{-}^{\mu
})+O(\epsilon )\,.  \label{fself0}
\end{equation}

Now that we eliminated the problematic $O(1/\epsilon )$ term, we can
safely take the limit $\epsilon \rightarrow 0$. It is at this limit where
we expect to obtain the universal expression for the self force. In this
limit all the $O(\epsilon )$ correction terms vanish, and we find
\begin{equation}
f_{self}^{\mu }=\frac{4}{3}q_{+}q_{-}(\dot{a}^{\mu }-a^{2}u^{\mu })+(\hat{f}
_{+}^{\mu }+\hat{f}_{-}^{\mu })\,.  \label{fself}
\end{equation}

As it stands, Eq. (\ref{fself}) provides a single relation for three
unknowns, $\hat{f}_{\pm }^{\mu }$ and $f_{self}^{\mu }$. In order to extract
from it the expression for $f_{self}^{\mu }$, we need to relate the latter
to the two partial self forces $\hat{f}_{\pm }^{\mu }$.
Since the self-force is the force that a charge experiences due to its own field,
 it must be proportional (for a prescribed worldline) to $q^2$,
where $q$ is the particle's charge.
In the limit of interest, $\epsilon \rightarrow 0$, the
trajectories of the two charges $\pm $, and also that of the dumbbell itself
(i.e. the representative point), all converge to the same worldline.
Therefore, the two partial self forces $\hat{f}_{\pm }^{\mu }$ will be given
by $\hat{f}_{\pm }^{\mu }=(q_{\pm }^{2}/q^{2})f_{self}^{\mu }$, where $%
q=q_{+}+q_{-}$ is the dumbbell's total charge. Substituting this in Eq. (\ref
{fself}), rewriting it as
\begin{equation}
\frac{4}{3}q_{+}q_{-}(\dot{a}^{\mu }-a^{2}u^{\mu })=f_{self}^{\mu }\,-(\hat{f%
}_{+}^{\mu }+\hat{f}_{-}^{\mu })=\left[ 1-\frac{q_{+}^{2}}{q^{2}}-\frac{
q_{-}^{2}}{q^{2}}\right] f_{self}^{\mu }\,\,,
\end{equation}
and noting that the term in squared brackets is nothing but $%
2q_{+}q_{-}/q^{2}$, we finally obtain the desired expression for the self
force:
\begin{equation}
f_{self}^{\mu }=\frac{2}{3}q^{2}(\dot{a}^{\mu }-a^{2}u^{\mu })\,\,.
\label{fsfinal}
\end{equation}
This agrees with Dirac's \cite{Dirac} expression (\ref{ALD}).

To summarize, let us formulate all elements of the above construction of $%
f_{self}^{\mu }$ by a single mathematical expression. This expression takes
the form
\begin{equation}
f_{self}^{\mu }=\frac{q^{2}}{2q_{+}q_{-}}\lim_{\epsilon \to 0}\left[
(1+\epsilon a_{w})f_{+}^{\mu }+(1-\epsilon a_{w})f_{-}^{\mu }+\frac{%
q_{+}q_{-}}{2\epsilon }a^{\mu }\right] \,.  \label{final}
\end{equation}
This involves the following manipulations, which are all justified (and
necessitated) by simple physical considerations: (i) the proper-time
weighting of the two mutual force (the factors $(1\pm \epsilon a_{w})$; (ii)
mass-renormalization (the last term in the squared brackets); (iii) the
inclusion of the partial self forces (the factor $q^{2}/2q_{+}q_{-}$); and
(iv) taking the limit $\epsilon \to 0$. This expression yields a universal,
orientation-independent, result, which conforms with the well known
expression (\ref{ALD}) for the self force.

Finally, we briefly discuss the justification of the (''bare'') equation of
motion $m_{bare}a^{\mu }=f^{\mu }$ in our case. We have {\em defined} the
total force $f^{\mu }$ as the proper-time derivative of the dumbbell's
non-electromagnetic energy-momentum $p^{\mu }$. Let us transform to a
Lorentz frame in which the dumbbell is momentarily at rest. In this frame
Eq.\ (\ref{defmom}) reads
\begin{equation}
p^{\mu }\equiv \int_{t=const}T_{(dumb)}^{\mu 0}d^{3}x^{i}\,,
\end{equation}
where $x^{i}$ denotes the three spatial Cartesian coordinates. For
simplicity let us approximate the dumbbell's stress-energy by that of a
continuous matter (plus, possibly, arbitrary number of point masses situated
at fixed locations on the dumbbell). Since the matter that composes each
element of the dumbbell is momentarily at rest, $T_{(dumb)}^{i0}$ vanishes,
and hence $p^{i}=0$. The dumbbell's energy in the rest frame is
\begin{equation}
p^{0}\equiv \int_{t=const}T_{(dumb)}^{00}d^{3}x^{i}\,.\qquad \qquad \text{
(rest frame)}
\end{equation}
This is by definition the dumbbell's bare mass. Thus, in the momentary rest
frame we have $p^{\mu }=(m_{bare},0,0,0)$. Rewriting this in a covariant
form (valid in any Lorentz frame), we obtain
\[
p^{\mu }=m_{bare}u^{\mu }\,.
\]
Since the dumbbell is approximated as rigid, its composition does not change
in time, hence $m_{bare}$ is time-independent. Differentiating now $p^{\mu }$
with respect to proper time, we obtain the desired equation of motion
\[
f^{\mu }=m_{bare}a^{\mu }\,.
\]
Recall that this is the ''bare'' equation of motion. After mass
renormalization, we obtain the equation of motion in its final, renormalized
form \cite{runaway}:
\begin{equation}
ma^{\mu }=f_{self}^{\mu }+f_{ext}^{\mu }=\frac{2}{3}q^{2}(\dot{a}^{\mu
}-a^{2}u^{\mu })+f_{ext}^{\mu }\,\,.  \label{eqmotion}
\end{equation}


\section{Extended Object With $N$ point charges}

\label{Nbody} In this section we shall consider a rigid extended
object with an arbitrary number $N$ of point charges located on it. The
charges are denoted $q_{i}$, where hereafter roman indices like $i,j,...$
run from $1$ to $N$. The total charge is $q=\sum_{i}q_{i}$. We shall
calculate the overall self force acting on the object, by a natural
extension of the method used above in the dumbbell case.


\subsection{Extended Object Kinematics}

We start by describing the extended object kinematics. We choose (quite
arbitrarily) a representative point inside this object and denote its
worldline by $z^{\mu }(\tau )$, and its four-velocity and four-acceleration
by $u^{\mu }\equiv dz^{\mu }/d\tau $ and $a^{\mu }\equiv du^{\mu }/d\tau $,
respectively, where $\tau $ is the proper time along this worldline.

The location of a charge $i$ at each moment $\tau $ is given by \footnote{%
Here and below there is no sum over repeated Latin indexes (such as i,j)
unless explicitly indicated.}
\begin{equation}
z_{i}^{\mu }(\tau )\equiv z^{\mu }(\tau )+\epsilon _{i}w_{i}^{\mu }(\tau
)\,\,,
\end{equation}
where $\epsilon _{i}\geq 0$ is the distance of the charge $i$ from the
representative point, and $w_{i}^{\mu }(\tau )$ is a unit spatial vector
normal to $u^{\mu }(\tau )$. We denote the proper time of this worldline by $%
\tau _{i}$ and its four-velocity and four-acceleration by $u_{i}^{\mu
}\equiv dz_{i}^{\mu }/d\tau _{i}$ and $a_{i}^{\mu }\equiv du_{i}^{\mu
}/d\tau _{i}$, respectively.

Since the object is rigid, and it moves in a non-rotational manner, the time
evolution of the spatial vectors $w_{i}^{\mu }$ is given by the Fermi-Walker
transport,
\begin{equation}
\dot{w}_{i}^{\mu }=(u^{\mu }a_{\nu }-u_{\nu }a^{\mu })w_{i}^{\nu }=u^{\mu
}a_{\nu }w_{i}^{\nu }\,\,\ .  \label{ob1}
\end{equation}
Repeating the above dumbbell kinematic calculations, we again find that
\begin{equation}
\frac{d\tau _{i}}{d\tau }=1+\epsilon _{i}a_{\mu }w_{i}^{\mu }\,\,\
\label{ob2}
\end{equation}
and
\begin{equation}
u_{i}^{\mu }=u^{\mu }\,.  \label{ob3}
\end{equation}
Again, the last equality implies that in the momentary rest frame of the
representative point, all charges are (momentarily) at rest too. One also
finds that
\begin{equation}
a_{i}^{\mu }=\frac{a^{\mu }}{1+\epsilon _{i}a_{\nu }w_{i}^{\nu }}\ .
\label{ob4}
\end{equation}

We shall be interested in the limit in which the object's size is taken to
be arbitrarily small, but its shape (including the location of the charges)
is unchanged in this limiting process. To describe this limit
mathematically, let $\epsilon >0$ denote the object's size, e.g. its
''radius'' (i.e. half the maximal distance between pairs of object's
points). We now define
\[
\epsilon _{i}\equiv \epsilon \alpha _{i}\,.
\]
The parameters $\alpha _{i}$ are thus dimensionless numbers of order unity
or smaller. The above limiting process is thus described by $\epsilon
\rightarrow 0$ with all parameters $\alpha _{i}$ kept fixed.

In the calculations below we shall make use of the results that were
obtained in Sec.\ {\ref{sec:dumbbell}} in the dumbbell case. Recall,
however, that in the latter case the representative point was chosen at half
the distance between the two charges. This cannot be done in the present
case (as long as $N>2$). In order to allow the implementation of the
dumbbell results to our case, we shall also need to consider, for each pair
of charges $i,j$, the worldline of the central point between these two
charges, which we denote $z_{ij}^{\mu }$:
\[
z_{ij}^{\mu }(\tau )\equiv \frac{1}{2}\left[ z_{i}^{\mu }(\tau )+z_{j}^{\mu
}(\tau )\right] =z^{\mu }(\tau )+\frac{1}{2}\left[ \epsilon _{i}w_{i}^{\mu
}+\epsilon _{j}w_{j}^{\mu }\right] \,.
\]
Obviously there exists a number $\epsilon _{ij}\geq 0$ and a unit vector $%
w_{ij}^{\mu }$ such that $\epsilon _{ij}w_{ij}^{\mu }=(1/2)(\epsilon
_{i}w_{i}^{\mu }+\epsilon _{j}w_{j}^{\mu })$. Then $\epsilon _{ij}$ is the
distance of this central point from the representative point, and (for $%
\epsilon _{ij}>0$) $w_{ij}^{\mu }$ is a vector normal to $u^{\alpha }$ which
satisfies the Fermi-Walker transport law, as one can easily verify. We
denote the proper time along the worldline $z_{ij}^{\mu }(\tau )$ by $\tau _{ij}$%
, and the four-velocity and four-acceleration by $u_{ij}^{\mu }\equiv
dz_{ij}^{\mu }/d\tau _{ij}$ and $a_{ij}^{\mu }\equiv du_{ij}^{\mu }/d\tau
_{ij}$, respectively. Obviously all the above kinematic relations satisfied
by the point charge $z_{i}^{\mu }(\tau )$, e.g. Eqs. (\ref{ob2}-\ref{ob4}),
are also satisfied by a central point $z_{ij}^{\mu }(\tau )$. Of particular
importance for the analysis below is the relation
\begin{equation}
a_{ij}^{\mu }\frac{d\tau _{ij}}{d\tau }=a^{\mu }\,,  \label{adt}
\end{equation}
which follows from Eqs. (\ref{ob2}) and (\ref{ob4}) (with ''$i$'' replaced
by ''$ij$'').

Let us finally emphasize that, for a particular pair $i,j$, the three points
$z_{i}^{\mu }$, $z_{j}^{\mu }$, and $z_{ij}^{\mu }$ satisfy all the
dumbbell's kinematic relations satisfied by the three dumbbell's points $%
z_{+}^{\mu }$, $z_{-}^{\mu }$, and $z^{\mu }$, correspondingly. This will
allow us to apply all the above dumbbell results to any pair $i,j$, though
with the dumbbell's central point $z^{\mu }$ replaced by $z_{ij}^{\mu }$
(and $\tau $ by $\tau _{ij}$, etc.). The dumbbell's length $\hat{\epsilon}
=2\epsilon $ is of course replaced by the distance between the charges $i$
and $j$, which we denote $\hat{\epsilon}_{ij}$.


\subsection{Calculation of the self Force}

To derive the self force acting on the extended object we shall use
energy-momentum considerations similar to those of Sec.\ \ref{sec:dumbbell}.
The four-momentum of the extended object $p^{\mu }(\tau )$ is defined just
as in the dumbbell case, by the integral (\ref{defmom}) over a hypersurface
of simultaneity. In analogy with Eq. (\ref{dpt}), we now have
\begin{equation}
dp^{\mu }=\sum_{i=1}^{N}dp_{i}^{\mu }+dp_{ext}^{\mu }\,,  \label{dpt2}
\end{equation}
where $dp_{i}^{\mu }$ denotes the contribution from all the electromagnetic
forces (sourced by all object's charges) acting on the $i$'th charge, and $%
dp_{ext}^{\mu }$ denotes the contribution from the overall external force.
The electromagnetic energy-momentum exchange with the charge $i$ is
\[
dp_{i}^{\mu }=f_{(em)i}^{\mu }d\tau _{i}\,,
\]
where $f_{(em)i}^{\mu }$ is the overall electromagnetic forces acting on the
charge $i$, given by

\[
f_{(em)i}^{\mu }=\hat{f}_{i}^{\mu }+
\sum^{N}_{\stackrel{\scriptstyle j=1}{j\neq i}}f_{j\rightarrow i}^{\mu}\,.
\]
Here $f_{j\rightarrow i}^{\mu }$ denotes the electromagnetic force that the
charge $j$ exerts on the charge $i$, and $\hat{f}_{i}^{\mu }$ denotes the
partial self force acting on this charge. Therefore,
\[
dp^{\mu }=dp_{ext}^{\mu }+\left( \sum_{i=1}^{N}\hat{f}_{i}^{\mu }d\tau
_{i}\right) +\left( \sum_{i=1}^{N}
\sum^{N}_{\stackrel{\scriptstyle j=1}{j\neq i}}f_{j\rightarrow i}^{\mu
}d\tau _{i}\right) \,.
\]
Since the external force is presumably well behaved as $\epsilon \rightarrow
0$ (it is essentially independent of $\epsilon $), we can use $dp_{ext}^{\mu
}\cong f_{ext}^{\mu }d\tau $, without bothering which proper time exactly
one should use. For the same reason we may replace $d\tau _{i}$ multiplying
the partial self force by $d\tau $. Defining the overall force acting on the
object to be
\[
f^{\mu }=\frac{dp^{\mu }}{d\tau }\,,
\]
we obtain
\begin{equation}
f^{\mu }\cong \left[ \sum_{i=1}^{N}\sum^{N}_{\stackrel{\scriptstyle j=1}{j\neq i}}\frac{d\tau _{i}}{d\tau }%
f_{j\rightarrow i}^{\mu }+\sum_{i=1}^{N}\hat{f}_{i}^{\mu }\right]
+f_{ext}^{\mu }\,.  \label{ftot3}
\end{equation}
The overall mutual force is the term including the double-sum over $i$ and $%
j $:
\begin{equation}
f_{mutual}^{\mu }=\sum_{i=1}^{N}\sum^{N}_{\stackrel{\scriptstyle j=1}{j\neq i}}\frac{d\tau _{i}}{d\tau }%
f_{j\rightarrow i}^{\mu }=\frac{1}{2}\sum_{i=1}^{N}\sum^{N}_{\stackrel{\scriptstyle j=1}{j\neq i}}\frac{d\tau
_{ij}}{d\tau }\left[ \frac{d\tau _{i}}{d\tau _{ij}}f_{j\rightarrow i}^{\mu }+%
\frac{d\tau _{j}}{d\tau _{ij}}f_{i\rightarrow j}^{\mu }\right] \,.
\label{fmut3}
\end{equation}
Consider the last term in squared brackets, for a particular pair of charges
$i,j$. This pair satisfies a ''dumbbell kinematics''; namely the kinematic
relations between the worldlines of the three points $z_{i}^{\mu }$, $%
z_{j}^{\mu }$, and $z_{ij}^{\mu }$ are exactly the same as those satisfied
by the three dumbbell's points $z_{+}^{\mu }$, $z_{-}^{\mu }$, and $z^{\mu }$%
, correspondingly. This allows us to apply the dumbbell's results to this
new two-charges system. In particular, Eqs. (\ref{dambmut0},\ref{dambmut})
now yield

\[
\frac{d\tau _{i}}{d\tau _{ij}}f_{j\rightarrow i}^{\mu }+\frac{d\tau _{j}}{
d\tau _{ij}}f_{i\rightarrow j}^{\mu }\cong -\frac{q_{i}q_{j}}{\hat{\epsilon}%
_{ij}}a_{ij}^{\mu }+\frac{4}{3}q_{i}q_{j}(\dot{a}_{ij}^{\mu
}-a_{ij}^{2}u^{\mu })\,,
\]
where $\hat{\epsilon}_{ij}$ is the distance between the two charges, and $%
\dot{a}_{ij}^{\mu }\equiv da_{ij}^{\mu }/d\tau _{ij}$. Note that $\hat{
\epsilon}_{ij}$, like all other object's distances, scales like $\epsilon $
(the object's size). Since the last term at the right-hand side is of order $%
\epsilon ^{0}$, we are allowed to replace $\tau _{ij}$ and $a_{ij}^{\mu }$
by the corresponding representative-point quantities, $\tau $ and $a^{\mu }$
(which we cannot do when treating the other term, the one proportional to $1/%
\hat{\epsilon}_{ij}$ ). With the aid of Eq. (\ref{adt}) we obtain
\begin{eqnarray*}
\frac{d\tau _{ij}}{d\tau }\left[ \frac{d\tau _{i}}{d\tau _{ij}}%
f_{j\rightarrow i}^{\mu }+\frac{d\tau _{j}}{d\tau _{ij}}f_{i\rightarrow
j}^{\mu }\right] &\cong &-\frac{q_{i}q_{j}}{\hat{\epsilon}_{ij}}\left(
a_{ij}^{\mu }\frac{d\tau _{ij}}{d\tau }\right) +\frac{4}{3}q_{i}q_{j}(\dot{a}%
^{\mu }-a^{2}u^{\mu }) \\
&=&-\frac{q_{i}q_{j}}{\hat{\epsilon}_{ij}}a^{\mu }+\frac{4}{3}q_{i}q_{j}(%
\dot{a}^{\mu }-a^{2}u^{\mu })\,.
\end{eqnarray*}
Notice that in the last expression all kinematic quantities are those
associated with the representative point, and the only reference to the two
charges is through $q_{i}$, $q_{j}$, and $\hat{\epsilon}_{ij}$. Substituting
this result back in Eq. (\ref{fmut3}) we obtain
\begin{equation}
f_{mutual}^{\mu }\cong -E_{es}a^{\mu }+\frac{2}{3}\left(
\sum_{i=1}^{N}\sum^{N}_{\stackrel{\scriptstyle j=1}{j\neq i}}q_{i}q_{j}\right) (\dot{a}^{\mu }-a^{2}u^{\mu
})\,,  \label{fbare3}
\end{equation}
where
\begin{equation}
E_{es}\equiv \frac{1}{2}\sum_{i=1}^{N}\sum^{N}_{\stackrel{\scriptstyle j=1}{j\neq i}}\frac{q_{i}q_{j}}{\hat{%
\epsilon}_{ij}}\,.  \label{selfe}
\end{equation}
This last expression is exactly the electrostatic energy of the system of $N$
charges (the factor $1/2$ corresponds to the fact that every pair $i,j$
appears twice in this sum).

The overall (bare) self force $f_{bare}^{\mu }$ is the term in squared
brackets in Eq. (\ref{ftot3}), which we write as
\begin{eqnarray}
f_{bare}^{\mu } &=&f_{mutual}^{\mu }+\sum_{i=1}^{N}\hat{f}_{i}^{\mu }
\nonumber \\
&=&-E_{es}a^{\mu }+\frac{2}{3}\left( \sum_{i=1}^{N}\sum^{N}_{\stackrel{\scriptstyle j=1}{j\neq i}}q_{i}q_{j}\right) (\dot{a}^{\mu }-a^{2}u^{\mu })+\sum_{i=1}^{N}\hat{f}%
_{i}^{\mu }+O(\epsilon )\,.  \label{fbare2}
\end{eqnarray}
Implementing now the mass-renormalization procedure,
given by Eq. (\ref{massren}),
and then taking the limit $\epsilon \rightarrow 0$, we obtain the
renormalized self force:
\begin{equation}
f_{self}^{\mu }=\frac{2}{3}\left( \sum_{i=1}^{N}\sum^{N}_{\stackrel{\scriptstyle j=1}{j\neq i}}q_{i}q_{j}\right) (\dot{a}^{\mu }-a^{2}u^{\mu })+\sum_{i=1}^{N}\hat{f}
_{i}^{\mu }\,.  \label{fself2}
\end{equation}

To factor out the partial self forces, we again use the fact that the self
force is quadratic in the charge, namely
\[
\hat{f}_{i}^{\mu }=(q_{i}^{2}/q^{2})f_{self}^{\mu }\,,
\]
where $q\equiv \sum_{i}q_{i}$ is the total charge. Transferring all partial
self forces to the left-hand side and then multiplying by $q^{2}$, we obtain
\[
\left[ q^{2}-\sum_{i=1}^{N}q_{i}^{2}\right] f_{self}^{\mu }=\frac{2}{3}%
q^{2}\left[ \sum_{i=1}^{N}\sum^{N}_{\stackrel{\scriptstyle j=1}{j\neq i}}q_{i}q_{j}\right] (\dot{a}^{\mu
}-a^{2}u^{\mu })\,.
\]
Noting that the two terms in squared brackets are equal, we obtain the self
force in its final form:
\begin{equation}
f_{self}^{\mu }=\frac{2}{3}q^{2}(\dot{a}^{\mu }-a^{2}u^{\mu })\,.
\label{fself3}
\end{equation}
The equation of motion is given by Eq. (\ref{eqmotion}), just as in the
dumbbell case.

\section{Continuously-charged extended object}
\label{continuous}
In this section we shall consider a rigid extended object which is
continuously charged. Again, we denote the object's size (e.g. its
''radius'') by $\epsilon $. Let $(X,Y,Z)$ be a system of comoving
Cartesian coordinates that parametrize the three-dimensional hypersurface of
simultaneity, and let $\bar{R}\equiv (X,Y,Z)$. The representative point (an
arbitrary point of the object) is taken to be e.g. at $\bar{R}=0.$ Note that
the worldline of any point of fixed $\bar{R}$ satisfies all the kinematic
relations described in the previous section. The charge distribution is
denoted $\rho (X,Y,Z)$. We assume that the charge distribution is fixed (in
the object's frame), i.e. $\rho (X,Y,Z)$ is independent of the proper time $%
\tau $.

The calculation of the self force proceeds in full analogy with the discrete
case discussed in the previous section, with the discrete charge $q_{i}$
replaced by the infinitesimal charge element $dq\equiv \rho dXdYdZ,$ and
with the summations replaced by integrals. There is a remarkable difference
between the two cases, though: In the discrete case, the demand for
consistency required us to take into account the partial self forces. No
such partial self forces appear in the continuous case (see below). This
makes the continuous case simpler and more elegant.

One can follow all the considerations and calculations of the previous
section, up to Eq. (\ref{fbare2}). In the continuous variant of this
equation, the double-sum becomes a double-integral:
\begin{equation}
\label{suminteg}
\sum_{i=1}^{N}\sum^N_{\stackrel{\scriptstyle j=1}{j\neq i}}q_{i}q_{j}\rightarrow \int \int \rho (\bar{R}%
_{1})\rho (\bar{R}_{2})d^{3}\bar{R}_{1}d^{3}\bar{R}_{2}=q^{2}\,,
\end{equation}
where $q\equiv \int \rho (\bar{R})d^{3}\bar{R}$ is the total charge. On the
other hand, the term including the partial self forces has only one
summation, so it would become a single integral. However, $\hat{f}_{i}^{\mu
} $ is proportional to $q_{i}^{2}$, and at the continuous limit this becomes
$(\rho dq)d^{3}\bar{R}$. That is, the ''integrand'' is proportional to $dq$,
which means that it actually vanishes at the infinitesimal limit. We
conclude that no partial self forces appear in the continuous limit. This
has a simple intuitive explanation: At the limit $N\rightarrow \infty $ in
which each charge is split into many smaller charges (such that the total
charge is conserved), the magnitude of the individual partial self forces
scales like $q_{i}^{2}\propto 1/N^{2}$, whereas their number only scales
like $N$. Therefore, the overall contribution of the partial self forces
scales like $1/N$ and hence vanishes at the continuous limit
\footnote{This property was also used in Eq. (\ref{suminteg}) where
we have replaced the double sum, which differs
from $q^2$, by a double integral--equal to $q^2$.
We use this property again below, when we replace
the double sum in Eq.\ (\ref{selfe}) by the double integral in Eq.\ (\ref{ees}).}.
(This is to be
contrasted with the situation of the mutual electromagnetic forces: The
magnitude of the mutual forces scales like $1/N^{2}$ too, but their number
scales like $N^{2}$, so the overall mutual force attains a non-vanishing
value at the limit $N\rightarrow \infty $.) The integral analog of Eq. (\ref
{fbare2}) is thus
\begin{equation}
f_{bare}^{\mu }=-E_{es}a^{\mu }+\frac{2}{3}q^{2}(\dot{a}^{\mu }-a^{2}u^{\mu
})+O(\epsilon )\,,  \label{contfin}
\end{equation}
where $E_{es}$ is the integral analog of Eq. (\ref{selfe}):
\begin{equation}
\label{ees}
E_{es}\equiv \frac{1}{2}\int \int \frac{\rho (\bar{R}_{1})\rho (\bar{R}_{2})%
}{|\bar{R}_{1}-\bar{R}_{2}|}d^{3}\bar{R}_{1}d^{3}\bar{R}_{2}\,.
\end{equation}
Note that $E_{es}$ is the electrostatic energy of the continuous charge
distribution.

The mass-renormalization (\ref{massren}) now removes the irregular term $%
E_{es}a^{\mu }$ in Eq. (\ref{contfin}), and (after taking the limit $%
\epsilon \rightarrow 0$) one arrives at the final expression for the self
force:
\begin{equation}
f_{self}^{\mu }=\frac{2}{3}(\dot{a}^{\mu }-a^{2}u^{\mu })\,.
\label{finalcont}
\end{equation}

\acknowledgments
This research was supported by The Israel Science
Foundation  (grant no. 74/02-11.1).

\newpage


\begin{thebibliography}{99}
\bibitem{relativistic}
In Ref. \cite{GriffithsOwen} the treatment is fully relativistic, except that the
accelerations of the dumbbell's two point charges are taken to be the same
as that of the central point, which is inconsistent with relativistic
kinematics. Despite of this, Griffiths and Owen's analysis yields the
correct expression for the term $c^{\mu}_{-1}/\epsilon$. [This may be
understood by comparing Eqs. (\ref{11}) and (\ref{12}) below]. In
Ref. \cite{Dillon},
it was argued that it is the non-relativistic treatment of the
four-acceleration that led to the mass-renormalization problem in Griffiths
and Owen's analysis. We disagree with this argument, because Griffiths and
Owen obtained the correct relativistic expression for the term
$c^{\mu}_{-1}/\epsilon$.
\bibitem{Review}
Reviews on the various aspects of the "self-force problem" can be found in
\cite{Pearle,Rohrlichbook,Erber,Teitelboimetal}.
\end{thebibliography}
\end{document}